\documentclass[aps,pre,showpacs,superscriptaddress,twocolumn]{revtex4}

\usepackage{graphicx}
\usepackage{amsmath}
\usepackage{amssymb}
\usepackage{bm}
\usepackage{times}
\usepackage{color}

\definecolor{yblue}{rgb}{0.06, 0.3, 0.57}
\usepackage[pdftex]{hyperref}
\hypersetup{colorlinks=true,linkcolor=yblue,citecolor=yblue,urlcolor=yblue}

\begin{document}

\title{Finite-size critical scaling in Ising spin glasses in the
mean-field regime}

\author{T.~Aspelmeier}
\affiliation{Felix Bernstein Inst.~Math.~Stat.~Biosci., G\"{o}ttingen,
Germany}
\affiliation{Univ. G\"{o}ttingen, Inst.~Math.~Stochast., D-37073
G\"{o}ttingen, Germany}
\affiliation{Max Planck Inst.~Biophys.~Chem., Stat.~Inverse Problems
Biophys.~Grp., D-37077 G\"{o}ttingen, Germany}

\author{Helmut G.~Katzgraber}
\affiliation{Department of Physics and Astronomy, Texas A\&M University,
College Station, Texas 77843-4242, USA}
\affiliation{Santa Fe Institute, 1399 Hyde Park Road, Santa Fe, New
Mexico 87501 USA}
\affiliation{Applied Mathematics Research Centre, Coventry University,
Coventry, CV1 5FB, England}

\author{Derek Larson}
\affiliation{Department of Physics, University of California, Santa
Cruz, California 95064, USA}

\author{M.~A.~Moore}
\affiliation{School of Physics and Astronomy, University of Manchester,
Manchester M13 9PL, United Kingdom}

\author{Matthew Wittmann}
\affiliation{Department of Physics, University of California, Santa
Cruz, California 95064, USA}

\author{Joonhyun Yeo}
\affiliation{Division of Quantum Phases and Devices, School of Physics,
Konkuk University, Seoul 143-701, Korea}

\date{\today}

\begin{abstract}

We study in Ising spin glasses the finite-size effects near the
spin-glass transition in zero field and at the de Almeida-Thouless
transition in a field  by Monte Carlo methods and by analytical
approximations. In zero field, the finite-size scaling function
associated with the spin-glass susceptibility of the
Sherrington-Kirkpatrick mean-field spin-glass model is of the same form
as that of one-dimensional spin-glass models with power-law long-range
interactions in the regime where they can be a proxy for the
Edwards-Anderson short-range spin-glass model above the upper critical
dimension. We also calculate a simple analytical approximation for the
spin-glass susceptibility crossover function. The behavior of the
spin-glass susceptibility near the de Almeida-Thouless transition line
has also been studied, but here we have only been able to obtain
analytically its behavior in the asymptotic limit above and below the
transition. We have also simulated the one-dimensional system in a field
in the non-mean-field regime to illustrate that when the Imry-Ma droplet
length scale  exceeds the system size  one can then be erroneously lead
to conclude that there is a de Almeida-Thouless transition even though
it is absent.

\end{abstract}


\maketitle

\section{Introduction}
\label{introduction}

The nature of the ordered state of spin glasses remains controversial,
despite decades of research. There are competing theories for the order
parameter of the low-temperature phase. The oldest is based on the
broken replica symmetry (RSB) theory of Parisi and
co-workers \cite{parisi:79,parisi:80,parisi:80a,parisi:83,mezard:84},
which gives the correct solution of the spin-glass problem in infinite
space dimensions (mean-field regime), that is, for the
Sherrington-Kirkpatrick (SK) model \cite{sherrington:75}. Alternative
theories based on scaling arguments include the droplet
model \cite{fisher:86,fisher:88,fisher:88b,bray:86,mcmillan:84b}. There
are also theories based on rigorous
calculations \cite{newman:92,newman:96,newman:98,newman:03,newman:07,stein:13}
which attempt to describe the behavior of these complex and
poorly-understood systems, yet contradict the mean-field picture of
Parisi. Recently, it has been argued that the RSB picture applies in
space dimensions $d >6$, while the droplet picture holds for $d \le 6$ 
\cite{moore:11,comment:parisi}. That $6$ might be the special dimension
down to which RSB might be applicable has been rigorously established
for a particular extreme choice of the spin-spin
interactions \cite{jackson:10}.

The thrust of the argument brought forward in Ref.~\onlinecite{moore:11}
concerns the phase transition which would take place in spin glasses in
an external field if there were RSB---the so-called de Almeida-Thouless
(AT) transition \cite{almeida:78}. Furthermore, it was argued in
Ref.~\onlinecite{moore:11} that when $d >6$ the AT transition line is
mean-field like so that $d_{\rm u} = 6$ is the upper critical dimension.
In renormalization group (RG) language this means its critical behavior
is controlled by a Gaussian fixed point. This point of view is supported
by the work of Castellana and Barbieri \cite{castellana:15}, who obtained
an equivalent result for a Dyson model on a hierarchical lattice.
However, the arguments of Ref.~\onlinecite{moore:11} and
\onlinecite{castellana:15} were based on {\em perturbative} results and
it has been recently suggested \cite{angelini:15} that there might be a
new non-Gaussian fixed point controlling the behavior in a field in high
space dimension. In addition, Castellana and Parisi \cite{castellana:15a}
further suggested on the basis of a numerical study of the Dyson
hierarchical model that a nonperturbative fixed point might also be
controlling the critical regime in the parameter range which corresponds
to $d \le 6$. We decided therefore to reexamine previously-published
Monte Carlo data in search of the nonpertubative fixed points.  Based on
our analysis, we conclude that at least for $d>6$ there is strong
evidence that the critical behavior both in a field and in zero field is
controlled by the trivial Gaussian fixed point. In addition, in
Sec.~\ref{ATlinecrit} below we argue that finite-size effects will
always make it difficult when $d \to 6^-$ to judge whether there is or
is not an AT line.

Monte Carlo simulations have of course been extensively used in an
attempt to understand the nature of spin glasses. Unfortunately in spin
glasses, even these state-of-the-art simulations are often plagued by
strong finite-size effects. In this paper we study in detail the form
which finite-size scaling (FSS) takes as this yields useful information
as to whether for $d > 6$ a nonperturbative fixed point or a Gaussian
fixed point is controlling the critical behavior.

The paper is structured as follows. In Sec.~\ref{model} we introduce the
models studied, as well as the measured observables and scaling
functions. In Sec.~\ref{universality} we study the universality of the
finite-size scaling function for the one-dimensional model with $\sigma
<2/3$ \cite{kotliar:83,katzgraber:03}, followed by a calculation of the
scaling function in Sec.~\ref{fscalingcalc}. Sections \ref{ATtransition}
and \ref{ATlinecrit} show results for  finite-size scaling at the AT
transition,  above and below the upper critical dimension, respectively.

\section{Model, Observables and Scaling Functions}
\label{model}

In practice, it is difficult to perform finite-size scaling studies on
large spin-glass systems when $d > 6$ because the number of sites in a
system of linear dimension $L$ increases very rapidly, as $L^d$, so that
the range of $L$ which can be studied is extremely limited. However, it
has been realized for some years now that a class of models in
one dimension with long-range interactions falling off with a power of
the distance between the spins can serve as a useful proxy for
short-range models in high
dimension \cite{kotliar:83,katzgraber:03,katzgraber:09b}. The
Hamiltonian of these power-law long-range models is given by
\begin{equation}
\mathcal{H}=-\sum_{ij} J_{ij} S_i S_j -\sum _i h_i S_i,
\label{Ham}
\end{equation}
where the sites $i=1, \cdots, N$ lie on a one-dimensional ring to
automatically enforce periodic boundary conditions.  The sum is over all
pairs of sites and the Ising spins $S_i \in \{\pm 1\}$ interact via
random couplings $J_{ij}$. The latter are independent random variables
of the form
\begin{equation}
J_{ij}= \epsilon_{ij}/R_{ij}^{\sigma},
\label{bonddef}
\end{equation}
where $\epsilon_{ij}$ is a random Gaussian variable with zero mean. It
is convenient to take the distance between spin $i$ and spin $j$,
$R_{ij}$, to be the chord distance between sites $i$ and $j$, so that
$R_{ij}=(N/\pi) \sin(\pi|i-j|/N)$. The variance of $\epsilon_{ij}$ is
fixed so that $(1/N)\sum_{i,j}J_{ij}^2=1$. The fields $h_i$ are drawn
from a Gaussian distribution with zero mean and variance $H^2$.  We
shall refer to the case when all the $h_i=0$ as the zero-field case.
Most of our simulational data have been obtained for this
one-dimensional proxy for the $d$ dimensional system in previous
numerical
studies \cite{katzgraber:03,katzgraber:03f,katzgraber:05c,katzgraber:08}.
Some of our data have also been obtained for {\it diluted} versions of
the models \cite{leuzzi:08,leuzzi:11}, where an average coordination
number $z_b = 6$ is chosen.  Details of these diluted models are also to
be found in Refs.~\onlinecite{katzgraber:09b} and
\onlinecite{larson:13}.

For $\sigma =0$, this model is the Sherrington-Kirkpatrick (SK)
model \cite{sherrington:75}, and for $0 < \sigma <1/2$ it shares the SK
universality class \cite{katzgraber:03,wittmann:12}. With our
normalization of the bonds, $T_c=1$ for all $\sigma <1/2$ when the field
$H=0$. Increasing $\sigma$ above $1/2$ is thought to be analogous to
changing an effective space dimension $d$ of a corresponding short-range
model. In the mean-field regime ($d > d_u=6$) the connection between
$\sigma$ and the equivalent space dimension $d$ is given
by \cite{katzgraber:03,katzgraber:09b,leuzzi:09,banos:12b}
\begin{equation}
d=\frac{2}{2 \sigma-1}.
\label{equivd}
\end{equation}
According to Eq.~(\ref{equivd}) our data for the case $\sigma=0.55$
therefore corresponds to working in an effective space dimension $d=20$.

We measure the wave-vector-dependent spin-glass susceptibility defined
by
\begin{equation}
\chi_{\rm SG}(k) = \! \frac{1}{L} \sum_{i, j} 
\! \left[\!
\Big( \! \langle S_i S_j\rangle \! - \!
\langle S_i \rangle \langle S_j\rangle \! \Big)^2 
\right]_{\rm av}
\!\!\!\!\!\! 
e^{ik\, (i-j)} .
\label{chidefSG}
\end{equation}
Note that we shall usually simply call $\chi_{\rm SG}(0)$ the spin-glass
susceptibility $\chi_{\rm SG}$. In Eq.~(\ref{chidefSG}) $\langle \cdots
\rangle$ represents a thermal average, whereas $[\cdots]_{\rm av}$
represents an average over the disorder.  The finite-size two-point
correlation length $\xi_L$ in a system of linear dimension $L$ is given
by \cite{ballesteros:00,katzgraber:06,larson:13}
\begin{equation}
\xi_L = \frac{1}{2 \sin (k_\mathrm{m}/2)}
\left[\frac{\chi_{\rm SG}(0)}{\chi_{\rm SG}(k_\mathrm{m})}
- 1\right]^{1/(2\sigma-1)} .
\label{eq:xiL}
\end{equation}
where $k_\mathrm{m} = 2 \pi / L$ is the smallest nonzero wave vector
compatible with the periodic boundary conditions. Note that for the
one-dimensional model, $L = N$, as $d = 1$, i.e., the linear size of the
system is the same as the number of spins $N$. These two quantities,
$\chi_{\rm SG}$ and $\xi_L$, are commonly studied in the spin-glass
literature, and it is the form of finite-size effects on these
quantities which is the subject of this paper.

The scaling form presented in
Refs.~\cite{jones:05,wittmann:14,wittmann:12} is different depending on
whether behavior is being controlled by a Gaussian fixed point or a
nontrivial fixed point. For example, if there is a nontrivial fixed
point controlling the critical behavior, the FSS form of the correlation
length $\xi_L$ in a system of $L^d$ spins takes the form
\begin{equation}
\xi_L/L=\tilde{\xi}\left[(T-T_c)L^{1/\nu}\right],
\label{NTfss}
\end{equation}
where the exponent $\nu$ is the exponent which describes the growth of
the correlation length in the infinite system, where  $\xi \sim
1/(T-T_c)^{\nu}$, and $\tilde{\xi}$ is the finite-size scaling
function. However, when the critical behavior is controlled by the
Gaussian fixed point, i.e., when one is above the upper critical
dimension, $d_u=6$ \cite{harris:76}, $\xi_L$ scales as
\begin{equation}
\xi_L/L^{d/d_u}=\tilde{\xi}\left[(T-T_c)L^{2 d/d_u}\right].
\label{Gfss}
\end{equation}
Thus, by finding which kind of FSS scaling works best, one can determine
the nature of the fixed point which controls the critical behavior.

To apply Eq.~(\ref{Gfss}) to the one-dimensional proxy model, we use
Eq.~(\ref{equivd}) for $d$ on the left of Eq.~(\ref{Gfss}), and on the
right side of the equation, we set $L^d=L \equiv N$ for
$d=1$ \cite{wittmann:14,wittmann:12}. Equation (\ref{Gfss}) therefore
becomes for $\sigma =0.55$
\begin{equation}
\xi_L/L^{1/[3 (2 \sigma-1)]} \to \xi_L/L^{10/3}=\tilde{\xi}\left[(T-T_c)L^{1/3}\right].
\label{scalingform}
\end{equation}

\begin{figure}
\includegraphics[width=\columnwidth]{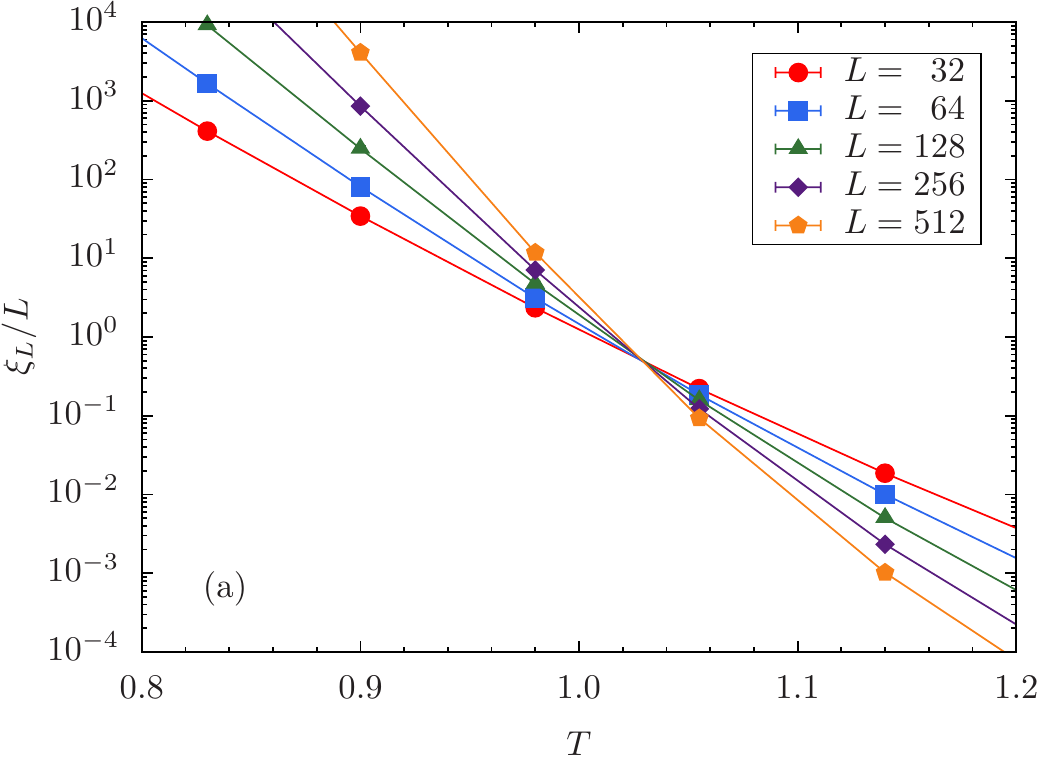}
\includegraphics[width=\columnwidth]{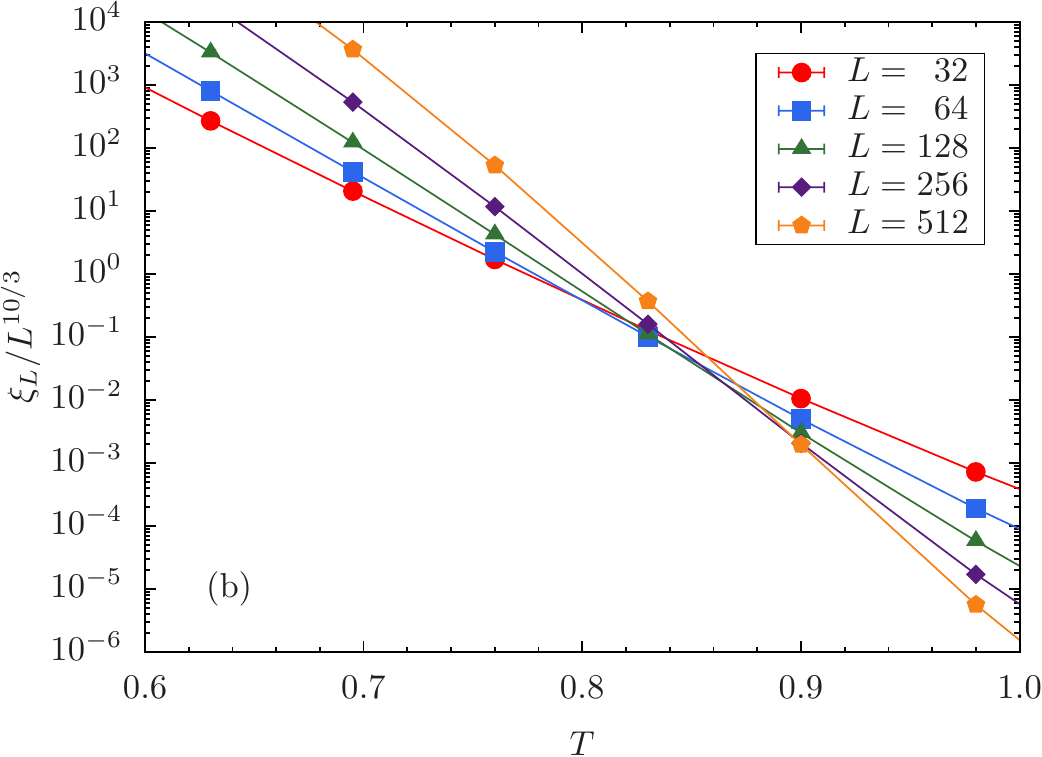}
\caption{(Color online) 
(a) Critical scaling form $\xi_L/L$ vs temperature $T$ for the
fully connected (complete) system with $\sigma =0.55$ in zero random
field $H$. (b) Mean-field scaling form $\xi_L/L$ vs
temperature $T$ for the fully connected (complete) system with $\sigma
=0.55$ in zero random field $H$.
}
\label{scalingconnxi}
\end{figure}

Figure \ref{scalingconnxi} shows the two scaling forms based on critical
scaling [Eq.~(\ref{NTfss})] and the mean-field scaling form expected
above the upper critical dimension [Eq.~(\ref{scalingform})].  We had
expected that the crossing of the curves for different $L$ values would
have been superior for the mean-field scaling form, but this is clearly
not the case for the studied system sizes.  A similar behavior when
searching for the AT line was found by Angelini and Biroli in
Ref.~\onlinecite{angelini:15} and they suggested as a consequence that
$d_u$ might not be $6$ for spin glasses in a field and that the critical
behavior for $d > 6$ might not be controlled by the Gaussian fixed point
but by some (as yet) undetermined nonperturbative fixed point.

\begin{figure}
\includegraphics[width=\columnwidth]{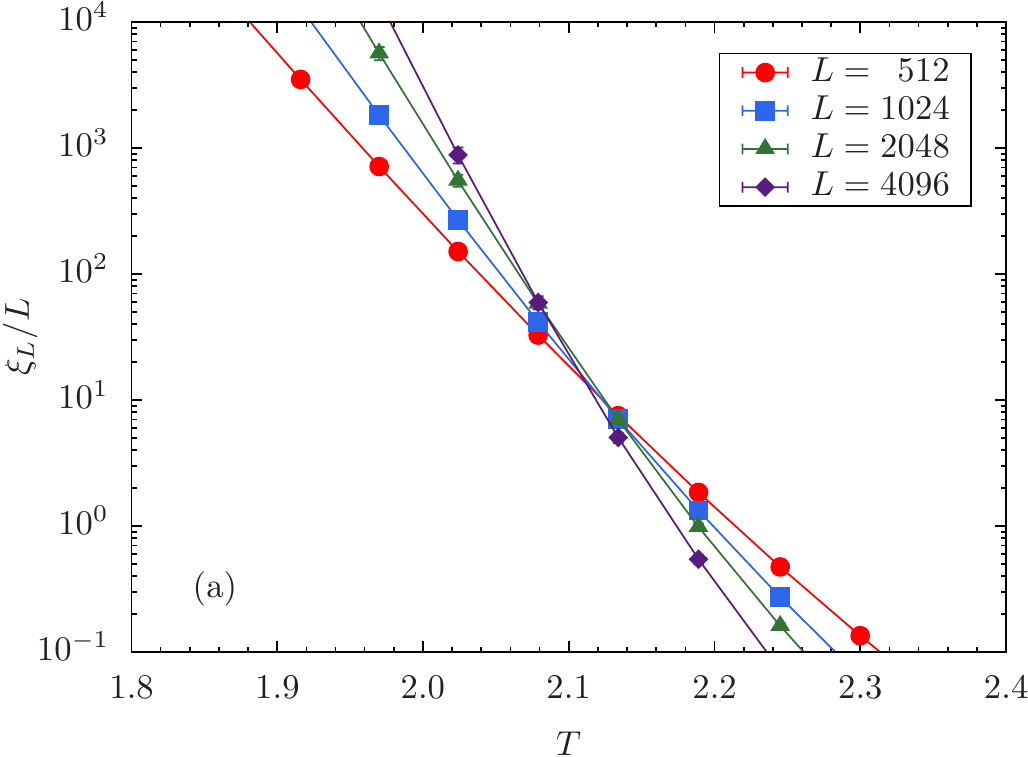}
\includegraphics[width=\columnwidth]{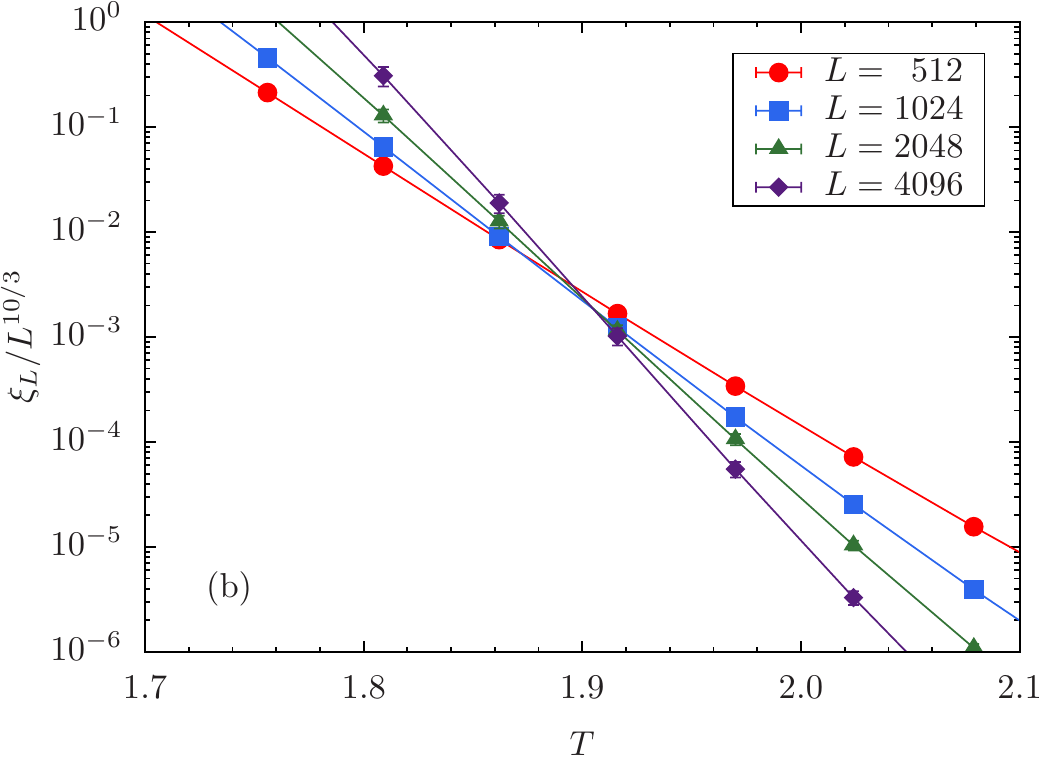}
\caption{(Color online) 
(a) Critical scaling form $\xi_L/L$ vs temperature $T$ for the
diluted model with $\sigma =0.55$ in zero random field $H$. (b) 
Mean-field scaling form $\xi_L/L$ vs temperature $T$ for the diluted
model with $\sigma =0.55$ in zero random field $H$.
}
\label{dilxi055crit}
\end{figure}

If one believes in the conventional wisdom that $6$ is the upper
critical dimension both in zero field and for the AT, then the only
possible explanation for the poor mean-field scaling is large
corrections to scaling in Fig.~\ref{scalingconnxi}.  On this
explanation, if one could obtain data for much larger systems than
$L=512$, then the crossing with mean-field scaling would eventually
become better than that for critical scaling. We cannot obtain such data
for the fully connected system, but we can for the diluted model and the
results for the two kinds of scaling functions are shown in
Fig.~\ref{dilxi055crit}.

There is some evidence that the crossing is indeed improving for the
mean-field scaling in these larger systems, but one could not really
argue that it is superior to the critical scaling form. Hence, using
these simple scaling plots we are unable to provide strong evidence for
$d_u = 6$.  Instead, we have to resort to an alternate approach to show
that mean-field scaling is the correct description of the critical
behavior.  Our approach is to analytically determine the scaling
function $\tilde{\xi}\left[(T-T_c) L^{1/3}\right]$ and show that the
simulational data fits well to this analytically calculated form.  We
find that it is possible to do this in zero field and we believe
that this is good evidence for the validity of mean-field scaling.  In a
field, finite-size effects are even larger in numerical work and on the
analytical side we have only been able to extract the asymptotic forms
for the scaling functions.

\begin{figure}
\includegraphics[width=0.97\columnwidth]{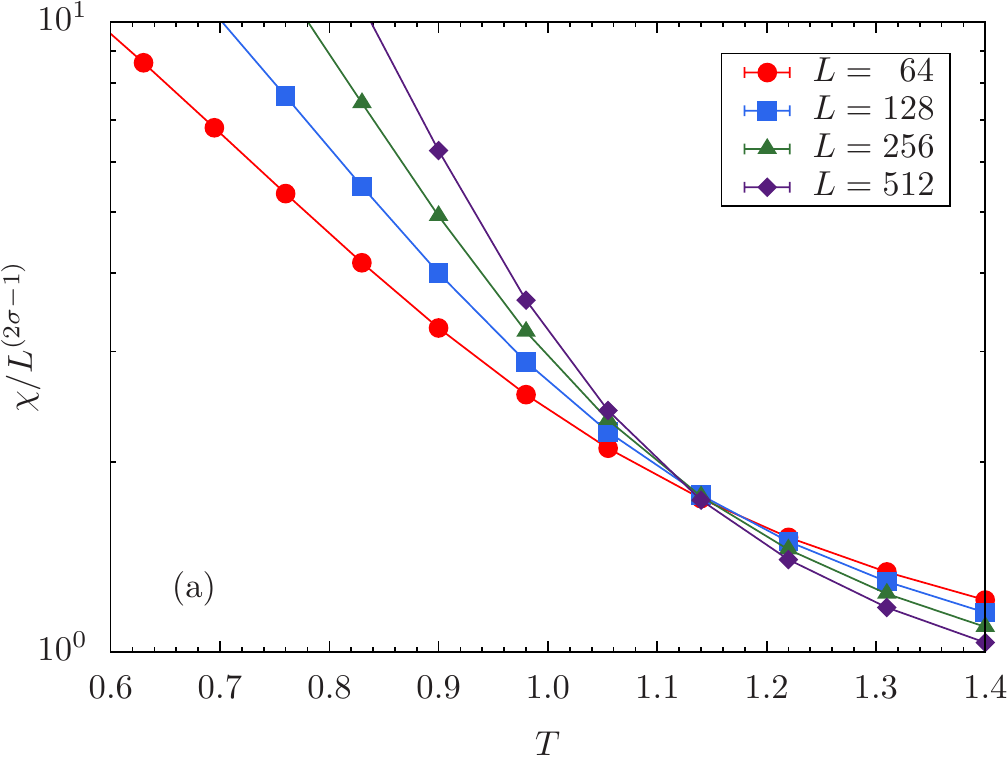}
\includegraphics[width=\columnwidth]{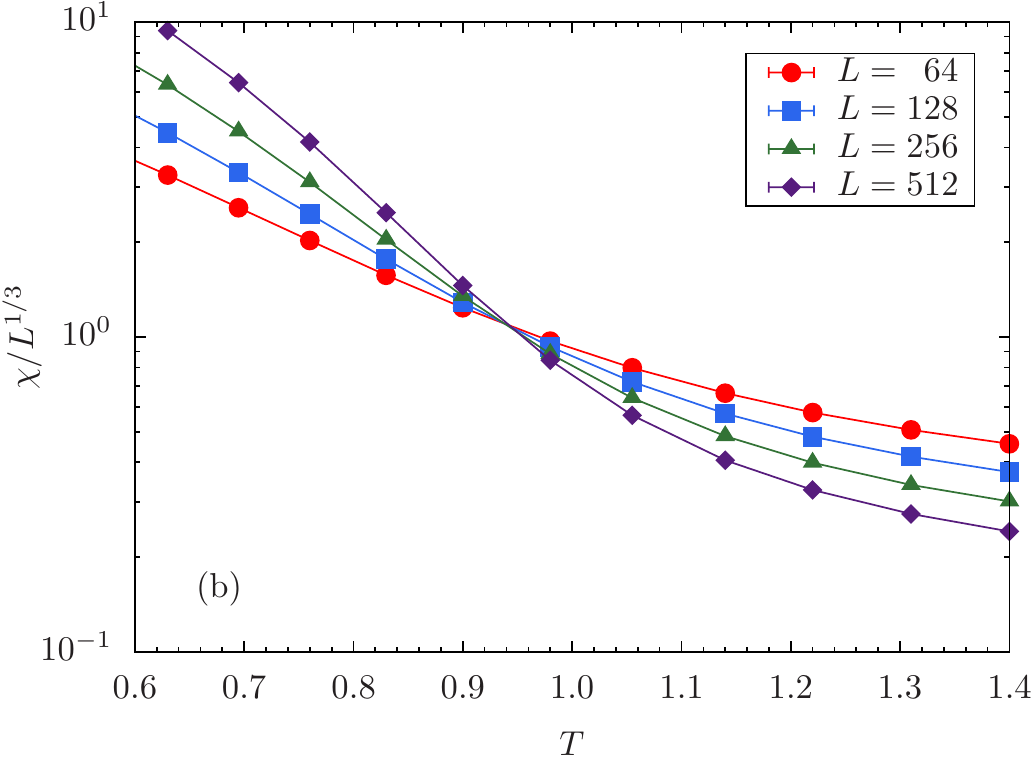}
\caption{(Color online) 
(a) Critical scaling form of the susceptibility $\chi/L^{2 \sigma-1}$
vs temperature $T$ for the fully connected (complete) system with
$\sigma =0.55$ when the random field $H=0$. (b) Mean-field scaling
form $\chi/L^{1/3}$ of the susceptibility $\chi/L^{2 \sigma-1}$ vs
temperature $T$ for the fully connected (complete) system with $\sigma
=0.55$ when the random field $H=0$. Note that $\chi \equiv \chi_{\rm
SG}$.
}
\label{chicrit}
\end{figure}

Rather than study the scaling function $\tilde{\xi}$, it is simpler to
study the equivalent scaling function for the spin-glass susceptibility
$\chi_{\rm SG}(0)$ obtained from the second moment of the spin-glass
order parameter $q$ where
\begin{equation}
q =\frac{1}{N} \sum_i S_i^{(1)} S_i^{(2)}.
\label{qdef}
\end{equation}
Here ``(1)'' and ``(2)'' refer to two independent copies of the system
with the same interactions $J_{ij}$. We have studied in particular the
second moment $q_2=[\langle q^2 \rangle]_{\rm av}$ and the quantity
\begin{equation}
\chi_{\rm SG}=N [\langle q^2 \rangle]_{\rm av}
\label{chisgdef}
\end{equation}
which is the spin-glass susceptibility in zero field. (Note that in a
finite system in zero field $\langle S_i \rangle=0$.) The analog of
the mean-field scaling form in Eq.~(\ref{scalingform})
is \cite{wittmann:12}
\begin{equation} 
\chi_{\rm SG}/L^{1/3}= \tilde{\chi}\left[(T-T_c) L^{1/3}\right].
\label{chiMFscaling}
\end{equation}
The analog of the critical scaling of Eq.~(\ref{NTfss}) is \cite{wittmann:12}
\begin{equation}
\chi_{\rm SG}/L^{2-\eta}=\tilde{\chi}\left[ (T-T_c)L^{1/\nu}\right],
\label{chicritscaling}
\end{equation}
where $2-\eta=2 \sigma-1$.  Again, $\tilde{\chi}(x)$ denotes the scaling
function, which will also be called $f(x)$.  The advantage of studying
$\chi_{\rm SG}$ rather than $\xi_L$ is that we can study it in the SK
universality class where $\sigma <1/2$, whereas $\xi_L$ is ill-defined
for these values of $\sigma$. The crossing of $\chi_{\rm SG}/L^{1/3}$
when plotted against the temperature $T$ for various values of the
system size $L$ were studied in Ref.~\onlinecite{wittmann:12} for
$\sigma =0$ and $0.25$. For $\sigma =0.55$ we present in
Fig.~\ref{chicrit} the corresponding scaling plots.

Notice that in the case of the susceptibility the quality of the
crossing is comparable for both the mean-field and critical scaling,
whereas for the correlation length the critical scaling form seemed
superior, at least for the fully connected system. However, the
temperature at which the curves cross provides an estimate of $T_c$, and
for both $\chi_{\rm SG}$ and $\xi_L$ critical scaling is indicating a
$T_c > 1$, whereas mean-field scaling indicates a $T_c < 1$. At the
level of mean-field theory the transition temperature would be $T_c =
1$, and the fluctuations about the mean field normally reduce the value
of the critical temperature $T_c$.  This clearly is an argument in
favor of using the mean-field scaling form. The same observation can be
made for the diluted model. For it the mean-field transition temperature
is $2.0564$ \cite{wittmann:12}, and the estimate of $T_c$ in
Fig.~\ref{dilxi055crit} for the case of $\sigma =0.55$ is certainly less
than this number using mean-field scaling, but larger than this for the
critical scaling form.

Standard finite-size scaling for mean-field scaling takes the 
form \cite{wittmann:12}
\begin{eqnarray}
\chi_{\rm SG}(T,L)&=&L^{1/3}\left[f(L^{1/3}t)+L^{-\omega}g(L^{1/3} t)+ \cdots\right]\nonumber\\
&+&d_0L^{2 \sigma-1}h(L^{1/3}t)+c_0+c_1t+ \cdots,
\label{fss}
\end{eqnarray}
where $t=T/T_c-1$, and the correction-to-scaling exponent is $\omega=2-3
\sigma$ \cite{kotliar:83}. In the limit $L \to \infty$ with $L^{1/3}t$
fixed, this equation reduces to the simpler form
\begin{equation}
\chi_{\rm SG}/L^{1/3}=f(L^{1/3}t)
\label{fssfunction}
\end{equation}
as then the corrections to scaling become negligible. In what follows,
we shall refer to the limit with $x=L^{1/3}t$ fixed as ``$L \to
\infty$'' as the finite-size scaling limit, and ``$f(x)$'' as the
finite-size scaling function for $\chi_{\rm SG}/L^{1/3}$.

In Sec.~\ref{universality} we outline the Br\'{e}zin and Zinn-Justin
procedure \cite{brezin:85} for calculating the universal scaling function
$f(x)$ for any space dimension $d > d_u=6$ (or $\sigma < 2/3$) and show
that our simulational data at $\sigma =0.0$, $0.25$, and $0.55$ are
consistent with being in the same universality class. In Sec.~\ref{fscalingcalc} we
determine $f(x)$ by using the mean-field equations of Thouless, Anderson
and Palmer (TAP) \cite{thouless:77}, as modified by Plefka
(TAPP) \cite{plefka:02}. We shall use in Sec.~\ref{ATtransition} these
same equations to determine the analog of the scaling function $f(x)$
at the AT transition in nonzero field, however only in the limit of
large $x$.  Finally, in Sec.~\ref{ATlinecrit} we discuss finite-size
problems which might  make one believe there is an AT line for $d \le
6$ ($\sigma \ge 2/3$) even though it is absent.

\section{Universality of the finite-size scaling function for $\sigma <2/3$}
\label{universality}

If the critical behavior is controlled by the Gaussian fixed point,
Br\'{e}zin and Zinn-Justin \cite{brezin:85} showed how the finite-size
scaling function $f(x)$ can, in principle, be calculated. The procedure
basically reduces to calculating the integral
\begin{equation}
\overline{Z^n}=\int dQ_{\alpha \beta} \exp\left[-F[\{Q_{\alpha\beta}\}]/kT\right],
\label{Zdef}
\end{equation}
where
\begin{eqnarray}
F[\{Q_{\alpha\beta}\}]/kT &=& \int d^dx\, \left[ \frac{1}{2}r
\sum_{\alpha<\beta}Q_{\alpha\beta}^2 
\right. \nonumber \\
&&\hspace*{-1.6cm} + \left.\frac{w}{6}
\sum_{\alpha<\beta<\gamma}Q_{\alpha\beta}
Q_{\beta\gamma}Q_{\gamma\alpha} + {\mathcal O}(Q^4)\right] .
\label{FQ}
\end{eqnarray}
The coefficient $r$ is essentially a measure of the distance from $T_c$,
i.e., it is related to the reduced temperature $t$.  The $Q^4$ terms are
irrelevant when calculating the scaling function, as are the usual
density gradient terms $(\nabla Q_{\alpha\beta})^2$ seen in such
free-energy functionals \cite{harris:76}, although they would have been
needed if we had tried to calculate the scaling function associated with
$\xi_L$.  $Q_{\alpha\beta}$ is related to the spin-glass order
parameter, and $\alpha$ takes the values $1$, $2$, $\cdots$, $n$, with
$n \to 0$. This integral should be adequate for calculating the
crossover scaling function $f(x)$ in the mean-field scaling regime,
i.e., for all $\sigma <2/3$. The form of the function is universal, and
the differences between fully connected spins or the diluted version of
the model, or the value of $\sigma$, just feed into the value of $T_c$,
the overall amplitude of $\chi_{\rm SG}/L^{1/3}$, and a multiplicative
factor associated with $t$.  For $\sigma >2/3$, when the behavior is not
controlled by the Gaussian fixed point but instead by the critical fixed
point \cite{harris:76}, the calculation of the scaling function is more
complicated. Its argument changes to $L^{1/\nu} t$ and the scaling
function is different from the universal form expected to apply for all
$\sigma <2/3$.

\begin{figure}
\includegraphics[width=\columnwidth]{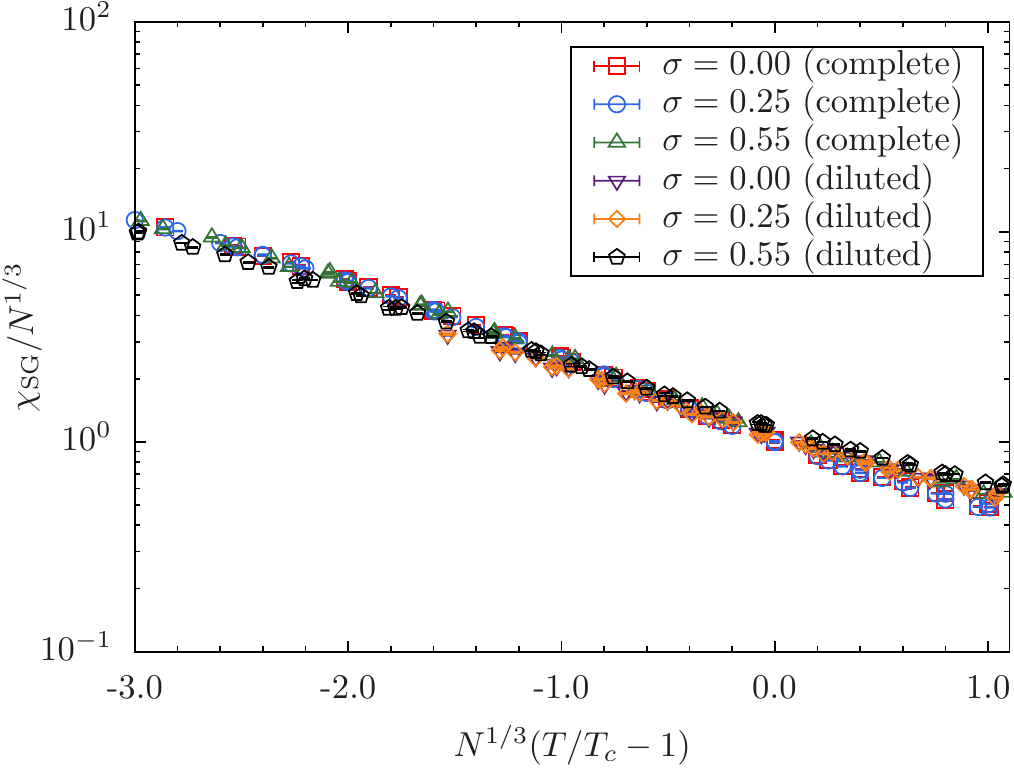}
\caption{(Color online) 
Reduced spin-glass susceptibility $\chi_{\rm SG}/N^{1/3}$ vs
$x=N^{1/3}(T/T_c-1)$, (recall, $N \equiv L$), i.e., the finite-size scaling
function $f(x)$ in zero field for all system sizes $N$ studied. For the SK
model ($\sigma = 0$) we simulated $N = 1024, \ldots, 4096$. For the diluted
model and $\sigma = 0$ we studied  $N = 2048, \ldots, 16384$.  For $\sigma =
0.25$ we studied $N = 512, \ldots, 4096$ for the complete (fully connected)
model and $N = 2048, \ldots, 16384$ for the diluted model. Data taken
from Ref.~\onlinecite{wittmann:12}. For $\sigma = 0.55$ we studied $N = 32,
\ldots, 512$ for the complete (fully connected) model and $N = 128, \ldots,
2048$ for the diluted model.  Data taken from Refs.~\onlinecite{katzgraber:03}
and \onlinecite{katzgraber:09b}.  For $\sigma=0.55$ for the complete
(fully connected) case we have taken $T_c \approx 0.94$, while for the diluted
case we use $T_c \approx 1.98$. Note the vertical logarithmic scale.
}
\label{scalingall}
\end{figure}

\begin{figure}
 \includegraphics[width=\columnwidth]{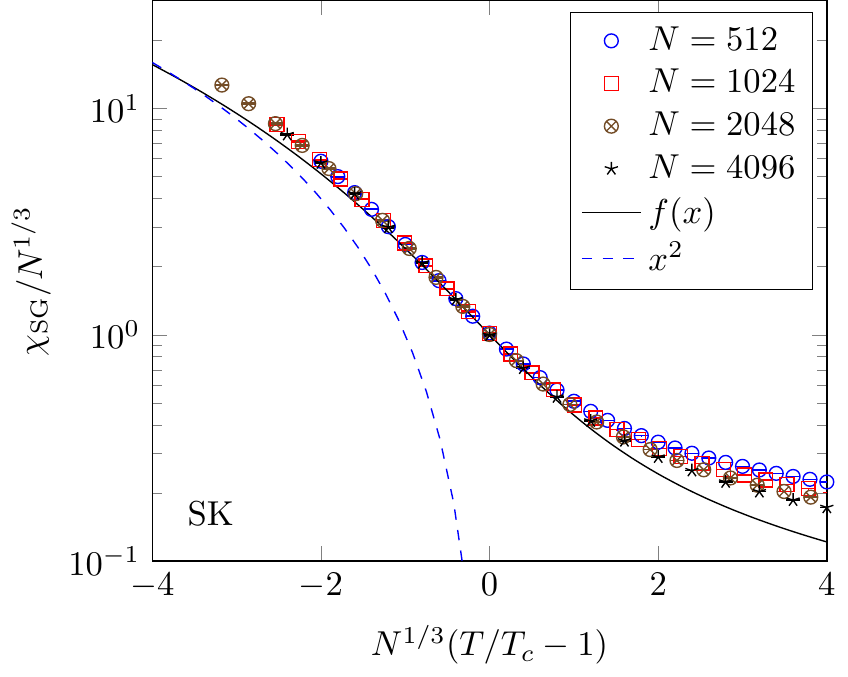}
\caption{(Color online) 
Reduced spin-glass susceptibility $\chi_{\rm SG}/N^{1/3}$ vs
$x=N^{1/3}(T/T_c-1)$, i.e., the finite-size scaling function $f(x)$ in zero
field for the SK model ($\sigma =0$). The data are taken from
Ref.~\onlinecite{wittmann:12}. For this model $T_c=1$ \cite{sherrington:75}. The
data for $x>1$ are strongly affected by finite-size effects. The solid curve
shows our approximation based on Eq.~(\ref{approxf(x)}) for the scaling
function $f(x)$ based on solving the TAPP equations. It gives, at large
positive $x$, $f(x) \to 1/(2x)$, while at large negative $x$, $f(x) \to x^2$.
The blue dashed curve is the asymptotic limit $f(x) \to x^2$ for negative $x$
values.
}
\label{scalingSK}
\end{figure}

In Fig.~\ref{scalingall} we plot results for $\chi_{\rm SG}/N^{1/3}$
vs the scaling variable $x=(T/T_c-1)N^{1/3}$ for $\sigma =0.0$,
$0.25$, and $0.55$ for both the fully connected (complete) model and for the
diluted model. The points include data for all the system sizes $N$
simulated (see caption). In the range $ 1>x>-3$ there is a fairly
satisfactory collapse of the data onto a single curve for the differing
values of $\sigma$ and for both the fully connected and dilute models.
None of the data have been linearly scaled on either the horizontal or
vertical axes of the figure, which would have been permissible while
staying in the same universality class. The data for $x>1$ are strongly
affected by finite-size effects, some of which can be seen in
Fig.~\ref{scalingSK}, which is why in Fig.~\ref{scalingall} we have
limited the horizontal range to $x < 1$.

Overall the data are consistent with a universal scaling function $f(x)$
for $\sigma <2/3$. If the behavior were controlled by a nonperturbative
fixed point rather than by the Gaussian fixed point, then such
universality of $f(x)$ would have to be understood.  Furthermore, as we
shall see in Sec.~\ref{fscalingcalc} below, it is possible to calculate
the function $f(x)$ explicitly.  Our results in Fig.~\ref{scalingSK}
turn out to be in satisfactory agreement with our approximation.

It is possible to determine the behavior of $f(x)$ as $x\to \pm \infty$
by simple arguments: when $x \to -\infty$, $\chi_{\rm SG} \to N q^2$,
and as $q \to -t$ in the scaling limit where $t \to 0$, that means that
$f(x) \to x^2$. The data in Fig.~\ref{scalingSK} are approaching this
estimate at large negative $x$. For $x \to \infty$, $\chi_{\rm SG}\to
1/(1-\beta^2)$ [see Eq.~\eqref{largeplusxh0}] for the SK model and also
from the TAPP equations, which implies that $f(x)\to 1/(2x)$ as
$\beta=1/(1+t)$. Again, the data shown in Fig.~\ref{scalingSK} seem to
be approaching this limit, but the finite-size effects are large for
positive $x$. This is not due to any inaccuracies in the TAPP equations,
but just points to the fact that in order to use the simplification
$\chi_{\rm SG}=1/(2t)$ [which leads to $f(x) \to 1/(2x)$] one needs to
work with rather small values of $t$.  However, at fixed large $x$, this
requires working with very large values of $N$, which are currently not
accessible numerically.

\section{Calculation of the scaling function $f(x)$}
\label{fscalingcalc}

In this section we outline how one can calculate the finite-size scaling
function $f(x)$. One approach would be to simply do the integrals in
Eq.~(\ref{Zdef}). Unfortunately, that is very difficult because of the
replica labels and the need to continue $n \to 0$. However, an approach
equivalent to this was used by three of us in Ref.~\onlinecite{yeo:05}
and it results in studying the finite-size scaling function for the
spherical SK spin-glass model, which according to the arguments in
the aforementioned reference should have an identical scaling function
$f(x)$. However, this approach is hard to extend to the behavior in a
field, so instead we present an approach which does permit, in
principle, an extension to finite fields.

Assuming that the scaling function $f(x)$ applies for all $\sigma <2/3$,
if we can calculate it for the SK model with $\sigma =0$ and that agrees
with data for (say) $\sigma =0.55$---as is the case in
Fig.~\ref{scalingall}---then the assumption would seem to be correct. To
calculate $f(x)$ for the SK model we use the TAP
equations \cite{thouless:77} as modified by Plefka \cite{plefka:02} and
refer to them as the TAPP equations. Plefka argued that in the presence
of an external field $h_i$ at each site $i$, the magnetization $m_i$ is
given by
\begin{equation}
m_i=\tanh \beta \big[h_i+\sum_j J_{ij} m_j -m_i \chi_\ell\big],
\label{PlefkaTAP}
\end{equation}
where the local susceptibility is given by 
\begin{equation}
\chi_\ell=N^{-1} \sum_i
\chi_{ii}=N^{-1}\sum_i \partial m_i /\partial h_i. 
\end{equation}
Plefka assumed that $\partial \chi_\ell/\partial m_i$ is of order
${\mathcal O}(N^{-1})$ and thus negligible when the inverse
susceptibility matrix is calculated from Eq.~(\ref{PlefkaTAP})
\begin{equation}
\chi_{ij}^{-1}=\delta_{ij}[\beta^{-1}(1-m_i^2)^{-1}+\chi_\ell]-J_{ij}.
\label{inversedef}
\end{equation}
Equations (\ref{PlefkaTAP}) and (\ref{inversedef}), with $\sum_j
\chi_{ij}\chi_{jk}^{-1}=\delta_{ik}$ form a closed set of equations for
the $m_i$ and $\chi_\ell$. They are not exact, unfortunately, as the
terms of ${\mathcal O}(N^{-1})$ can, for certain quantities, combine to
make ${\mathcal O}(1)$ contributions \cite{aspelmeier:04}. We believe
that such possibilities are unimportant in our calculation of $f(x)$.
Our argument for this is that the use of these equations gives in our
finite-size scaling limit the same results as can be obtained by the
spherical model SK spin glass mapping \cite{yeo:05}, which we think is
exact in zero field. For zero fields, in our scaling regime, $m_i \to
0$, and Eq.~(\ref{inversedef}) simplifies to
\begin{equation}
\chi_{ij}^{-1}=\delta_{ij}[\beta^{-1}+\chi_\ell]-J_{ij}.
\label{inversedefhi0}
\end{equation}
The self-consistency equation for $\chi_\ell$ is then conveniently
written in terms of $z=\beta^{-1} +\chi_\ell$ as
\begin{equation}
z-\beta^{-1}=N^{-1}\sum _i \frac{1}{z-\lambda_i},
\label{selfconlamb}
\end{equation}
where $\lambda_i$ are the eigenvalues of the matrix $J_{ij}$. The
physical solution is the solution which has the largest real value of
$z$.

In the large-$N$ limit, the $N$ real eigenvalues $\lambda_i$ are
described by the semicircle distribution with support between $-2$ and
$2$. Then Eq.~(\ref{selfconlamb}) reduces to
\begin{equation}
z-\beta^{-1}=1+\frac{(z-2)-\sqrt{(z-2)(z+2)}}{2}, 
\label{selfconlargeN}
\end{equation}
which gives $z=\beta+\beta^{-1}$. We want to calculate
\begin{eqnarray}
\chi_{\rm SG} &\equiv& \frac{1}{\beta^2} N^{-1} \sum_{i,j} \chi_{ij}^2 \nonumber\\
&=& \frac{1}{\beta^2} N^{-1}\sum_i \frac{1}{(z-\lambda_i)^2}.
\label{chiSGlargeN}
\end{eqnarray}
In the large-$N$ limit, the sum can be done and gives
\begin{equation}
\frac{1}{2\beta^2} \cdot \frac{z}{\sqrt{z^2-4}-1} ,
\end{equation}
which reduces to 
\begin{equation}
 \chi_{\rm SG} \to \frac{1}{1-\beta^2}, 
\label{largeplusxh0}
\end{equation}
on substituting $z =\beta+\beta^{-1}$. It is this result which we use to 
determine the limit of $f(x)$ as $x \to \infty$ (see Fig.~\ref{FSShall}). 

In principle, for finite $N$ values, one could solve for $z$ numerically
using Eq.~(\ref{selfconlamb}). However, this is difficult for large $N$.
Instead, we give an approximate solution which seems in practice to be
quite accurate. As $x \to -\infty$, $z \to \lambda_{\rm max}$ and throughout
the scaling region differs from $\lambda_{\rm max}$ by terms of ${\mathcal
O}(1/N^{2/3})$. The largest eigenvalue itself has the form
$\lambda_{\rm max}=2+{\mathcal O}(1/N^{2/3})$. Let us introduce the variable
$u=(z-\lambda_{\rm max})N^{2/3} > 0$ and the notation
$\Delta=(\lambda_{\rm max}-\lambda_1)N^{2/3}$, where $\lambda_1$ is the next
largest eigenvalue. Then we separate off the first two terms in the sum
in Eq.~(\ref{selfconlamb}) and approximate the rest by
Eq.~({\ref{selfconlargeN}) after replacing $(z-2)$ by
$(z-\lambda_{\rm max})$ \cite{yeo:05}. The left-hand side of
Eq.~(\ref{selfconlamb}) becomes
\begin{equation}
z-\beta^{-1} \to 1-x/N^{1/3} +{\mathcal O}(1/N^{2/3}).
\label{LHSselfcon}
\end{equation}
The right-hand side becomes
\begin{equation}
\frac{1}{N^{1/3} u}+\frac{1}{N^{1/3}(u+\Delta)}+1-\sqrt{u/N^{2/3}}
+{\mathcal O}(1/N^{2/3}) .
\label{RHSselfcon}
\end{equation} 
Thus, correct to ${\mathcal O}(1/N^{2/3}),$ we have as our basic
approximation for $u$,
\begin{equation}
-x=\frac{1}{u}+\frac{1}{u+\Delta}-\sqrt{u}.
\label{uequation}
\end{equation}
Within the same approximation, the sample with gap $\Delta$ gives for $f(x)$
\begin{equation}
f(x)= \frac{1}{u^2}+\frac{1}{(u+\Delta)^2}+\frac{1}{2 \sqrt{u}}.
\label{approxf(x)}
\end{equation}
To calculate the bond-averaged value of $f(x)$ we must average over the
spacing $\Delta$ which we do with the Wigner surmise distribution for
it \cite{rodgers:89}.

Before comparing with the numerical data we need to introduce the
pseudocritical temperature $T_c(N)$ \cite{billoire:11a,castellana:11}.
If one studies the function $-f^{\prime}(x)/f(x)$, it has a peak at
$T_c$. However, in a system of finite size $N$, this peak is shifted to
$T_c(N)$, where in the mean-field regime,
\begin{equation} 
T_c(N)=T_c-\frac{a}{N^{1/3}}.
\label{pseudodef}
\end{equation}
For the SK model $T_c = 1$ and typical values for $a$ are $\sim 0.2$,
but this depends on the function being studied \cite{billoire:11a}. When
trying to construct the universal scaling function $f(x)$ for different
models it is natural to shift the horizontal axis so that the peaks for
the different models coincide at $x=0$, which can by done by redefining
$x$ so that $x=[T/T_c(N)-1]N^{1/3}$. This definition of $x$ differs from
the old definition by $a+{\mathcal O}(1/N^{1/3})$. Thus, when comparing
to our numerical data, one can shift the curves by an amount $a$ to
improve the fit, and this is what we did in Fig.~\ref{scalingSK}. With
this shift, the overall agreement is quite satisfactory, considering the
simplicity of the approximation. We suspect that it might be possible to
calculate $f(x)$ exactly, but that remains a challenge for the future.

\section{Finite-size scaling at the Almeida-Thouless transition}
\label{ATtransition}

In this section we shall discuss finite-size scaling at the AT
transition \cite{almeida:78}. The upper critical dimension of the AT line
is expected to be the same as in zero field, that is,
$d_u=6$ \cite{bray:80b}. For the long-range model, that translates to
$\sigma =2/3$. Note that in a field $\langle S_i \rangle $ is nonzero,
and we have to study the cumulant second moment, i.e.,
\begin{equation}
\chi_{\rm SG}=\frac{1}{N} \sum_{i,j} 
\big[ \langle S_i S_j \rangle-\langle S_i \rangle \langle S_j 
\rangle \big]_{\rm av}^2 .
\label{chidefh}
\end{equation}
In a field we only have numerical data for the one-dimensional
long-range model with $\sigma =0.55$. In Fig.~\ref{FSShall} we show the
mean-field scaling form $\chi_{\rm SG}/N^{1/3} = f_H(x)$ against
$x=N^{1/3}[T/T_c(H) -1]$. The finite-size effects are strongly visible
on the low-temperature side of the transition.

\begin{figure}
\includegraphics[width=\columnwidth]{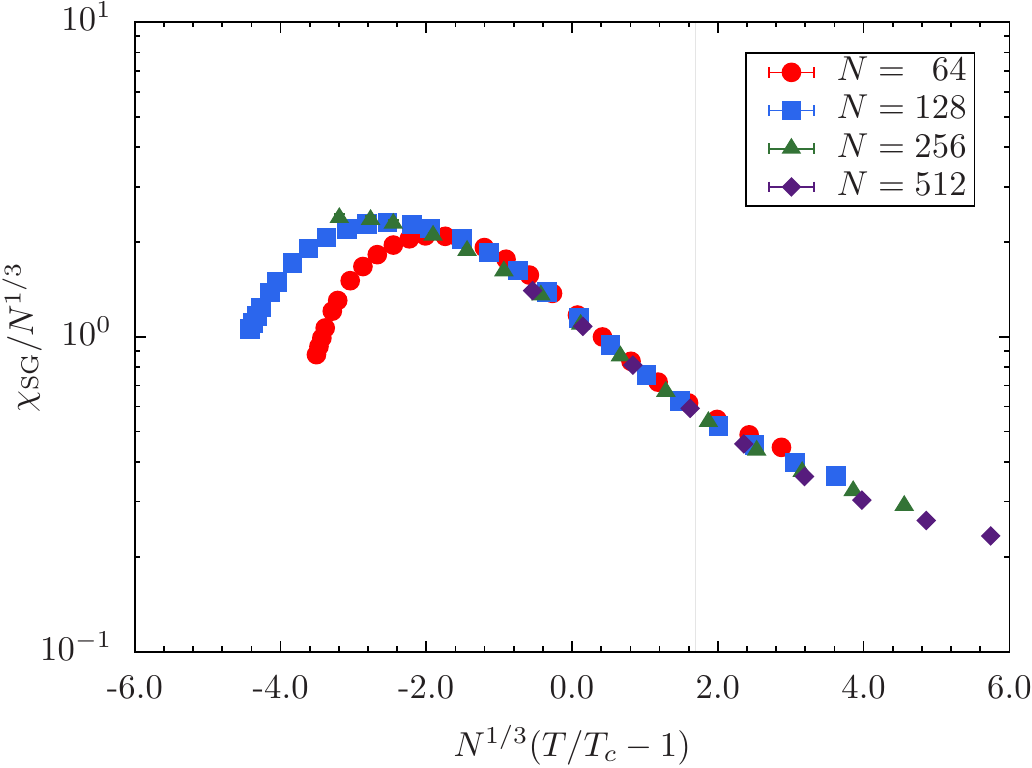}
\caption{(Color online)
Reduced spin-glass susceptibility $\chi_{\rm SG}/N^{1/3}$ vs
$x=N^{1/3}[T/T_c(H)-1]$, that is the finite-size scaling function
$f_H(x)$ in a random field of standard deviation $0.10$ when
$\sigma=0.55$.  For this model $T_c(H=0.1) \approx 0.815$.}
\label{FSShall}
\end{figure}
We now turn to understanding the form of the finite-size scaling
function $f_H(x)$ near the AT transition. At the formal level, the
analog of Eq.~(\ref{FQ}) for the AT transition involves just the
fields in the replicon sector $\tilde{Q}_{\alpha \beta}$, which are such
that $\sum_{\beta}\tilde{Q}_{\alpha \beta}=0$ \cite{bray:80b}. The
replicated partition function is
\begin{equation}
\overline{Z^n}=\int d\tilde{Q}_{\alpha \beta} \exp\left[-F[\{\tilde{Q}_{\alpha\beta}\}]/kT\right],
\label{Zdefh}
\end{equation}
where the effective functional is
\begin{eqnarray}
&F[\{\tilde{Q}_{\alpha\beta}\}]/kT& = \int d^dx\, \left[\frac{1}{4}\tilde{r}
\sum \tilde{Q}_{\alpha\beta}^2
 \right. \nonumber \\ 
&+& \hspace*{-1.0cm} \left. \frac{w_1}{6}
\sum \tilde{Q}_{\alpha\beta}
\tilde{Q}_{\beta\gamma}\tilde{Q}_{\gamma\alpha} 
+\frac{w_2}{6}\sum \tilde{Q}_{\alpha\beta}^3\right]. 
\label{FBR}
\end{eqnarray}
Here the convention has been adopted that the sums over replica indices
are unrestricted. Note that $\tilde{Q}_{\alpha \alpha}=0$. At the AT
line, $\tilde{r} =0$ in the mean-field approximation and the two
couplings $w_1$ and $w_2$ both depend on the field $H$. We would expect
as a consequence the finite-size scaling function $f_H(x)$ to depend on
both $x$ and the strength of the field $H$. It is because the effective
field theory is a cubic field theory that the upper critical dimension
$d_u = 6$ for the AT line \cite{bray:80b}. Unfortunately, the integrals
in Eq.~(\ref{FBR}) needed to calculate $f_H(x)$ are even more difficult
to do than those of the zero-field case and other methods have to be
used to understand the finite-size scaling function $f_H(x)$.

We have tried solving the TAPP equations for the SK limit in the
presence of a field numerically.  We obtained the solution for a given
bond and field realization at high temperatures, and followed the
solution down to lower temperatures, for $N$ values up to $400$. At
temperatures well above $T_c(H)$, we obtained values for $\chi_{\rm SG}$
consistent with those in Fig.~\ref{FSShall}.  At large positive values
of $x$ one is effectively in the regime where one can use the locator
expansion \cite{bray:79} on the TAPP equations. The
result \cite{plefka:02} is that Eq.~(\ref{largeplusxh0}) is generalized
to
\begin{equation}
\chi_{\rm SG} \to \frac{\chi_{\rm SG}^0}{1-\beta^2 \chi_{\rm SG}^0},
\label{largeplusx}
\end{equation}
where \cite{sharma:10}
\begin{equation}
\chi_{\rm SG}^0 = \frac{1}{N}\sum_i(1-m_i^2)^2 . 
\end{equation}
We stress that this result holds just for the large-$N$ limit and $T>T_c(H)$. For the SK model in a random field $H$, $\chi_{\rm SG}^0$ can be
determined explicitly. At the AT transition $T_c(H)$,
$\chi_{\rm SG}^0 = T_c(H)^2$, and so for $T$ close to $T_c(H)$,
\begin{equation}
\chi_{\rm SG} \to 
\frac{T_c(H)^2}{[1-\beta T_c(H)][1+ \beta T_c(H)]} \to \frac{1}{2t}T_c(H)^2 ,
\label{highxh}
\end{equation}
where $t =[T/T_c(H)-1]$. For $H =0.1$, $T_c(0.1) = 0.819428$ for the SK
model, so $\chi_{\rm SG} \to 0.33573/t$. Unfortunately, the data in
Fig.~\ref{FSShall} have not been obtained at large enough values of $N$
(here the largest $N$ value is $ 512$) or small enough values of $t$, to
see this behavior clearly. However, the calculated values of $\chi_{\rm
SG}$ are consistent with the result presented in Eq.~(\ref{largeplusx}).

As the temperature is reduced to well below $T_c(H)$, the solution of
the TAPP equations in the large-$N$ limit is expected to reduce to
$\chi_{\rm SG} \to 1/|t|$ \cite{plefka:02}. For the $N$ values for which
we could obtain solutions, i.e., $N < 400$, this behavior was not
visible. In fact, most samples showed a peak in $\chi_{\rm SG}$ well
above $T_c(H)$, followed by a fall at lower temperatures. We suspect
that the fall at low temperatures visible in Fig.~\ref{FSShall} might by
connected with the fall seen in the TAPP equations. The decrease in
$\chi_{\rm SG}/N^{1/3}$ at large negative $x$ values seen in
Fig.~\ref{FSShall} is clearly a finite-size effect.

We suspect that in the absence of finite-size effects $f_H(x)$ would
actually continue to grow $\propto x^2$ at large negative $x$, due to
replica symmetry breaking effects \cite{dedominicis:05}, and not follow
the expectations based on the solution of the TAPP equations, which
would be that $f(x) \to 1/|x|$. In Ref.~\onlinecite{dedominicis:05} it
was shown that for the SK model, where the Parisi RSB broken order
parameter is $q(x)$,
\begin{equation}
\chi_{\rm SG}= \frac{N}{3}
\left\{
\int_0^1 q^2(x) dx - \left[\int_0^1 q(x) dx \,\right]^2\right\}.
\label{RSBeffect}
\end{equation}
At large negative $x$ one would therefore expect that because of these
replica symmetry breaking effects that $f_H(x) \to B x^2$ so that
$\chi_{\rm SG}$ is proportional to $N$. Using the results in
Refs.~\onlinecite{dedominicis:05} and \onlinecite{thouless:80} for
$q(x)$ in a field, one can calculate the coefficient $B$ and it is of
order $q_{EA}$ on the AT line, which is small ($\approx 0.2)$ when
$H=0.1$. However, the data in Fig.~\ref{FSShall} at negative $x$ values
are not extensive enough to provide a clear verification of these
predictions.

\section{Numerical searches for the Almeida-Thouless line when 
$d \le 6$ ($\sigma \ge 2/3$)}
\label{ATlinecrit}

In Sec.~\ref{introduction} we stated that there is likely no AT line
when $d \le 6$. As a consequence, we were surprised when Castellana and
Parisi \cite{castellana:15a} recently claimed that in the Dyson
hierarchical model numerical evidence suggested the existence of an AT
line at $\sigma =0.68 > 2/3$ which corresponds to an effective space
dimension $d <6$. At the transition they reported values for the
critical exponents which were not close to their mean-field values,
which lead them to suggest that the behavior was being controlled by a
nonperturbative fixed point.

In this section we discuss a problem which arises when trying to
determine the existence of the AT line in dimensions where there might
be no AT line. It is again a finite-size problem. If there is no AT line
and the droplet picture applies, then the correlation length $\xi_D$ in
the system is the Imry-Ma length \cite{imry:75}, determined by equating
the free energy cost of flipping a region of size $\xi_D$, $ k
T_c(\xi_D/\xi) ^{\theta}$ to the energy which might be gained from the
random applied field, which is $\sqrt{q} H \xi_D^{d/2}$ (see for
example, Ref.~\onlinecite{yeo:15}).  Here $\xi$ denotes the zero-field
correlation length $\sim 1/|t|^{\nu}$.  In our one-dimensional model, $\theta
=1-\sigma$ \cite{moore:12}. Then
\begin{equation}
\frac{\xi_D}{\xi} \sim \left[\frac{H_{\rm AT}}{ H }\right]^{2/(2 \sigma-1)},
\label{ImryMa}
\end{equation}
where 
\begin{equation}
H_{\rm AT} \equiv kT_c |t|^{(\gamma+\beta)/2} , 
\end{equation}
which is the scaling expectation for the form of the AT
line \cite{yeo:15}, should it exist.  At the borderline value of $\sigma
=2/3$, $\xi_D$ grows rapidly for small fields $\propto 1/H^6$. In order
to see droplet behavior one requires system sizes $L > \xi_D$.
Otherwise, one might be tempted to think there is an AT line. For
$\sigma = 0.75$ we plot $\xi_L/L$ as a function of the field $H$ in
Fig.~\ref{larsonfig}. Here, $\xi_D$ grows at small $H$ $\propto 1/H^4$.
Figure \ref{larsonfig} shows that the droplet model prediction that
$\xi_L \sim \xi_D$ fails when $\xi_L > L$, as then finite-size effects
are clearly making $\xi_L$ deviate away from $\xi_D$.  The basic message
is that to see droplet model behavior one needs to study system sizes $L
> \xi_D$.  When studying fields where $\xi_D> L$, one can be misled into
thinking there is evidence for an AT line, as discussed at great length
in Ref.~\onlinecite{larson:13}.  We suspect this is why the authors of
Ref.~\onlinecite{castellana:15a} thought there was an AT line at $\sigma
=0.68$. In fact, the growth of $\xi_D$ as $1/H^{6}$ when $\sigma \to
2/3^+$ will always make it very difficult to obtain data for the regime
where $L > \xi_D$.

\begin{figure}
\includegraphics[width=\columnwidth]{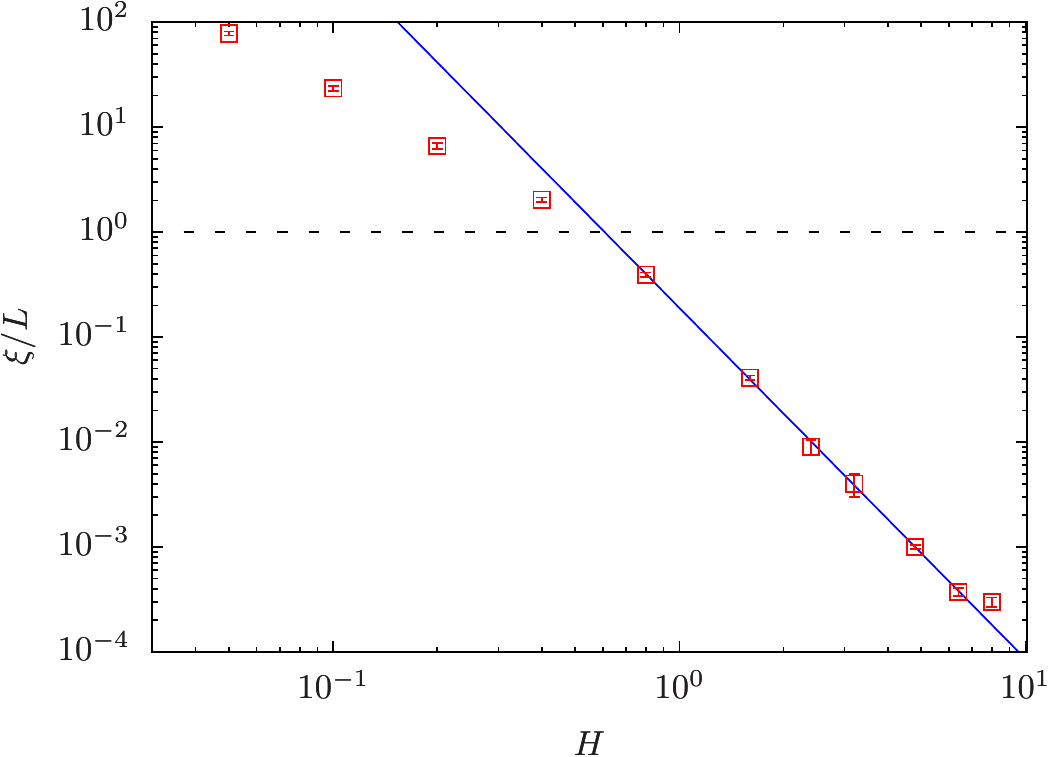}
\caption{(Color online) 
Correlation length $\xi_L/L$ over a large range of field values for $L =
1024$, $T = 0.48$, and $\sigma =0.75$. The horizontal dashed line is a guide
to the eye marking the point where $\xi_L = L$. A change in behavior
for $\xi < L$ is apparent. The solid (blue) line marks the regime where
the Imry-Ma argument \cite{imry:75} is valid.
}
\label{larsonfig}
\end{figure}

\section{Conclusions}

We have studied finite-size effects on critical scaling in Ising spin
glasses both in zero field and finite field in the regime where
mean-field scaling is expected. We believe that the conventional wisdom
that both types of transition have $6$ as the upper critical dimension
is supported by the numerical data gathered from previous studies, even
though strong finite-size effects are present.  For the zero-field case,
we have found a simple approximation for the crossover function for the
spin-glass susceptibility. The finite-field case is far more difficult,
but we have been able to determine the asymptotic form of the crossover
function by allowing for the non-self-averaging features of the Parisi
order parameter $q(x)$ which occur below the AT transition.

We should point out that there is a lack of self averaging generally
throughout the critical scaling regime. Thus, in zero field, we have
studied in the SK limit the distribution function of $\chi_{\rm SG}$ at
$T=T_c$ which arises from different realizations of the bonds $J_{ij}$
that has a well-defined distribution.  The zero-field problem seems
sufficiently simple such that one day the scaling function $f(x)$ might
be determined analytically; as a by-product one might then obtain the
corresponding distribution functions.

We have argued previously that evidence for an AT line when $d < 6$
might be just a consequence of not allowing for the effects of
finite-size effects. In order to see the droplet picture emerging
clearly, one needs the linear system size $L$ to be larger than the
Imry-Ma length $\xi_D$. However, this length scale can be very long at
the fields commonly used in most numerical studies. This means that when
$L \le \xi_D$ one can easily be mislead into believing that there is a
transition in a field. For example, from the data presented in
Fig.~\ref{larsonfig} for the one-dimensional model with $\sigma=0.75$,
one needs system sizes $L$ larger than $1024$ sites, as well as fields
stronger than $H_R \approx 0.7$ to see the droplet behavior.  Our hope
is that future studies first verify the needed system sizes $L > \xi_D$
before claiming the existence of a spin-glass state in a field.

\acknowledgments  

We thank Peter Young for participating in the initial stages of this
work and providing extremely useful feedback on this project.
Furthermore, we thank Satya Majumdar and Yan Fyodorov for discussions on
the possibility of calculating the scaling function $f(x)$ exactly, as
well as Leo Alcorn for providing help with GraphClick
\cite{comment:graphclick} (used to digitize the data in
Fig.~\ref{larsonfig}). J.Y.~was supported by Basic Science Research
Program through the National Research Foundation of Korea (NRF) funded
by the Ministry of Education (Grant No.~2014R1A1A2053362). M.W.~was
supported through the National Science Foundation (Grant
No.~DMR-1207036).  H.G.K.~acknowledges support from the National Science
Foundation (Grant No.~DMR-1151387) and would like to thank Hitachino
Nest for inspiration. H.~G.~K.'s research is based upon work supported
in part by the Office of the Director of National Intelligence (ODNI),
Intelligence Advanced Research Projects Activity (IARPA), via MIT
Lincoln Laboratory Air Force Contract No.~FA8721-05-C-0002. The views
and conclusions contained herein are those of the authors and should not
be interpreted as necessarily representing the official policies or
endorsements, either expressed or implied, of ODNI, IARPA, or the
U.S.~Government. The U.S.~Government is authorized to reproduce and
distribute reprints for Governmental purpose notwithstanding any
copyright annotation thereon.  We thank ETH Zurich for access to their
Brutus cluster, the Texas Advanced Computing Center (TACC) at The
University of Texas at Austin for providing HPC resources (Stampede
cluster), and Texas A\&M University for access to their Ada, Curie, Eos
and Lonestar clusters.

\bibliography{refs,comments}

\begin{thebibliography}{59}
\bibitem[{com({\natexlab{a}})}]{comment:parisi}
\bibinfo{note}{Criticisms of this argument \cite{parisi:12} were further
  discussed in a very recent paper by two of us \cite{yeo:15}.}
\end{thebibliography}

\end{document}